# Sub-wavelength Lithography of Complex 2D and 3D Nanostructures without Dyes

**Raghvendra P Chaudhary, Arun Jaiswal, Govind Ummethala, Suyog R Hawal, Sumit Saxena and Shobha Shukla***

*Nanostructures Engineering and Modeling Laboratory, Department of Metallurgical Engineering and Materials Science, Indian Institute of Technology Bombay, Mumbai, MH, India 400076*

*sshukla@iitb.ac.in

**Abstract:** One-photon or two photon absorption by dye molecules in photopolymers enable direct 2D & 3D lithography of micro/nano structures with high spatial resolution and can be used effectively in fabricating artificially structured nanomaterials. However, the major bottleneck in unleashing the potential of this useful technique is the indispensable usage of dyes that are extremely expensive, highly toxic and usually insoluble in commercially available photopolymers. Here we report a simple, inexpensive and one-step technique for direct-writing of micro/nanostructures, with sub-wavelength resolution at extremely high speeds without using any one photon or two photon absorbing dye. We incorporated large amount (20 weight %) of inexpensive photoinitiator into the photopolymer and utilized its two-photon absorbing property for sub-wavelength patterning. Complex 2D and 3D patterns were fabricated with sub-micron resolution, in commercially available liquid photopolymer to show the impact/versatility of this technique.

Keywords: Two-photonn Lithography, femtosecond laser, phtoinitiator, 2D/3D micro/nanostructures, nonlinear optics, direct writing

# 1. Introduction:

Miniaturization of optical and electronic devices is a daunting task as it requires use of micro/nanolithographic techniques for fabrication of sub-wavelength 2D/3D nanostructures. The current state of the art facilities involve the use of electron beam lithography (EBL) or focused ion beam lithography (FIBL) [1],[2],[3] for fabrication of such 2D sub-micro/nano structures. These techniques are however very expensive, non-scalable, time consuming and limited towards fabrication of only planar sub-micro/nano structures with low throughput. Layer by layer assembly and self-assembly are other methods for fabricating 3D nanostructures but offer limited choice of materials and involve high level of complex chemistry. Fabrication of 2D/3D nanostructures via a single step can be achieved using direct two photon laser (TPL) lithography which is scalable, comparatively inexpensive and a vacuum less process [4],[5],[6],[7][8],[9],[10]. Since two-photon absorption is confined to the focal volume of the tightly focused high intensity infrared (IR) laser beam, it provides 3D control over photochemical and photophysical excitations, in the vicinity of the material. Thus materials can be artificially engineered to achieve potential optical response which is beyond the reach of natural homogeneous materials[11],[12],[13],[14] and can be utilized in the rapidly advancing field of nanophotonics, flexible electronics, plasmonics and metamaterials [15],[16],[17],[18],[19],[20],[21].

Typically, TPL is enabled by absorbing two photons simultaneously by dye molecules, which transfers energy to catalytic photoinitiator to start the polymerization of monomer/oligomer. Though a lot of two photon absorbing dyes with very high two photon absorption (TPA) cross sections have been developed in the past, commercial unavailability and high cost limits their potential applicability[22]. Various attempts to use direct laser lithography without use of any dye on commercial photopolymers/resins have been made sporadically but resulted in structures significantly larger than the wavelength of laser used. The main culprit for increased line-width is the small two photon absorption coefficient (~1GM) of photoinitiator present in the photopolymer, in the IR region (800nm-1000nm) as compared to IR absorbing dyes ($10^3$-$10^5$GM). The basic requirement for sub-wavelength writing without dye, is the presence of a sensitive and efficient photoinitiator that requires low optical dose with short exposure time for generating radicals in a very small area[23]. Commercially available photoinitiators such as Irgacure 369 and Lucirin TPOL have been explored towards this goal but found to be less efficient due to their small TPA cross-sections, as mentioned before. Lucrin TPO-L an acylphosphine oxide radical photoinitiator has however been found to be a better photoinitiator for TPL with several advantages such as easy solubility in most of the commercial resins and availability in solid as well as liquid form making it suitable candidate for easy processing and higher loading. Although it has a small TPA cross section (< 1.2GM), structures fabricated with this chemistry have shown excellent resolution and integrity at relatively low laser powers owing to high radical quantum yield of 0.99[24]. Baldacchini et.al used TPO-L as a photoinitiator in sartomer resin formulation back in year 2004 for two photon writing. A few weight percentage of photoinitiator yielded polymerized features as small as 5 μm[25]. Since then few more groups have tried to incorporate TPO-L photoinitiator with small loading concentrations (~3 weight %) for the fabrication of polymeric micro-cantilevers, optically active microstructures, structures doped with gold nanoparticles and biological applications achieving feature sizes in the vicinity of 4μm[26],[27],[28],[29]. Very recently, Matthias et al fabricated branched hollow fiber microstructures with predefined circular pores using Ormocer, an organic – inorganic hybrid resin material, pre-loaded with Lucirin TPO-L by employing average laser power of 105 mW and a writing speed of 5 mm/s[30]. Most of these studies have resulted in extremely large feature sizes. In the pursuit of developing a dye-less technique for fabricating sub-wavelength feature sizes, we explore effects of using higher loading of photoinitiator on fabricated structure. In this study, we have shown that 2D and 3D nano-structures with resolutions as good as 140 nm at high writing speeds of 100 mm/s can be achieved with 20 weight% loading of TPO-L. Extensive characterization and optimization has been performed to validate the finding. By employing high writing speed, fabrication of nanostructures in polymer matrix can be performed, which is extremely useful for photonic & optoelectronics devices[18].This advancement will enable the fabrication of real 2D & 3D photonic materials with low cost and high-throughput.

2. **Experimental Procedure:**

Two liquid resins, triacryclic monomers, tris(2-hydroxy ethyl) isocyanurate triacrylate (SR368) and ethoxylated (6) trimethylolpropane triacrylate (SR499) were procured from Sartomer. SR499 reduces structural shrinkage whereas SR368 confers hardness to the structure during laser irradiation. Ethyl-2,4,6-Trimethylbenzoylphenylphosphinate (Lucirin-TPOL) which is a photoinitiator, was procured from BASF. Composition of the liquid resins SR368 and SR499 in the ratio of 50:50 by weight is used for photopolymerization. 20 weight% Lucirin-TPOL was added to the above composition which causes free radical polymerization during irradiation with laser. No two photon absorption dye was used at any stage during the fabrication of polymeric structures using femtosecond laser. A thin film of the respective mixture was spin coated on a cover glass. Prior to spin coating, the cover glass was silanized with 3-aminopropyl triethoxysilane to facilitate the adhesion of the mixed resin. The samples were irradiated with 140fs, 800nm laser pulses from Ti-sapphire laser oscillator operating at a repetition rate of 80MHz (Coherent Chameleon, Ultra I). The laser was coupled via an acousto-optic modulator to an inverted microscope (Olympus-IX81) as shown in figure1. The laser beam was focused inside the resin using a 100X objective with numerical aperture of 0.9. Acrylate-coated cover glass was mounted on a XYZ piezo-stage (PI-nanopositioner E-725) of the inverted optical microscope. The lithographic process was monitored on an EMCCD coupled to the microscope. This lithography setup was controlled using an in-house developed code using LabVIEW. Writing speed was varied from 100µm/s to 100mm/s. After writing, the unexposed resin was washed off in dimethylformamide (DMF) for SEM imaging. Since the temperature rise in the focal volume is sufficient enough for annealing [31] no post baking of the sample was performed.

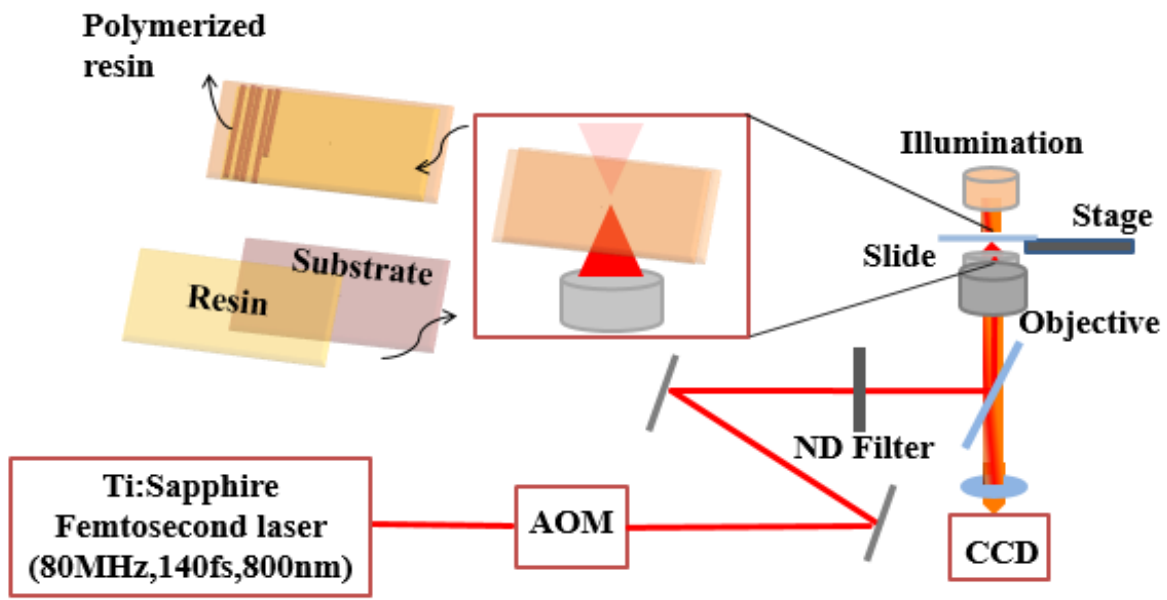

*Figure 1:- Schematic of the two-photon lithography setup for fabrication of complex 2D and 3D nanostructures.*

3. **Results and Discussion:-**

Microscopic analysis of the fabricated structures was performed using scanning electron microscopy (SEM). Figure 2 shows sample images of sub-wavelength 2D nano structures patterned in pure polymer loaded with 20 weight % TPO-L. Sub-wavelength feature sizes of ~ 140 nm (figure 2a) were observed for an irradiation power of 0.17W at writing speed of 5000μm/s. Thus we have demonstrated that sub-wavelength resolution can be achieved without using two photon dyes even with high speed, for inexpensive prototyping. We have proved the capability of this method to write complex and mixed structures of sub-wavelength resolution. The effects of laser power on feature size in pure resin shown in figure 2(b) were investigated by varying the laser power at fixed writing speed. Polymeric linewidths were found to be decreasing with decreasing laser power. Though structures could be written at powers less than 0.15 watt (as seen in the microscope CCD during writing) they were not strong and got washed off during washing step. Minimum power at which structures were sturdy enough to hold shape even during washing step has been termed as polymerization threshold power. Power *vs* linewidth trend can be attributed to enhanced interaction of photons per unit time per unit volume of the material with increasing power, giving rise to free radical polymerization in larger area. It is apparent from figure 2b that even though the photoinitiator has less TPA cross-section, it is able to absorb significant amount of laser power at 800nm and start the polymerization process. It should also be noted that even though the required laser power in this process is significantly higher than the traditional two photon dye loaded technique where the required laser power is only few mW, we are still able to achieve state of the art resolution with higher power at high photoinitiator loading.

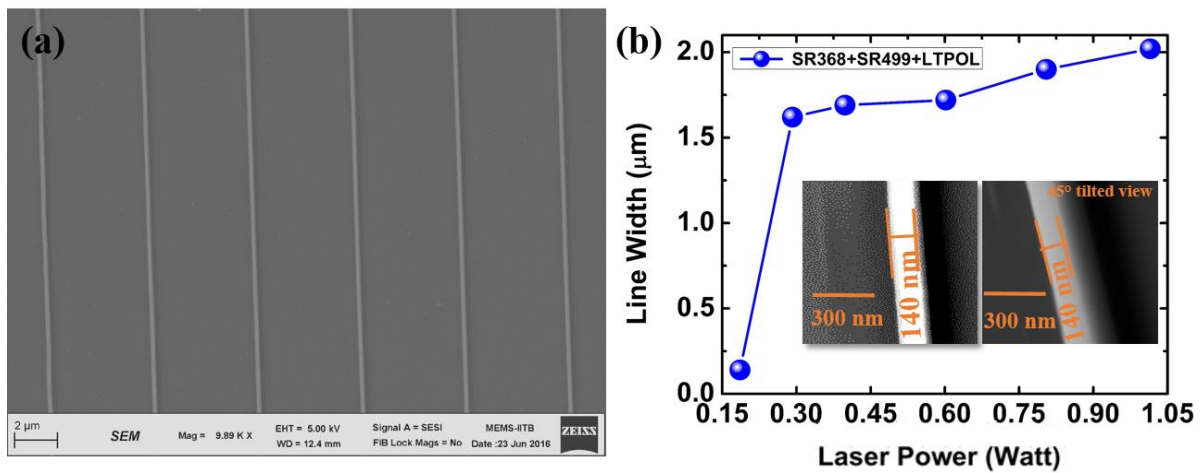

*Figure 2 - (a) SEM image of sub-wavelength (~140nm) parallel lines written using femto second laser at 800 nm, in polymeric resin at a speed of 5000μm/s and 0.17W power. The separation between consecutive lines is 5μm. (b) Laser power vs line width measurement performed at constant scanning speed. The operating wavelength of Laser used is 800 nm wavelength. Inset in 2(b) shows the zoomed in images of the top view and 45 degree tilted view of a single line respectively.*

Figure 3(a) shows the SEM image of 3D mesh structure of six layers written in pure resin at 0.25W power at a scanning speed of 5000μm/s. Figure 3(b) shows a zoomed in image of the structure in figure 3(a), showing the line width of ~ 1μm with layers for this 3D mesh pattern. We found that the each layer consists of three constitutive layer like patterns as marked by the yellow box in figure 3(b). As we are writing the structures from top to bottom, these consecutive layers may form due to the interference between incident laser beam and the laser beam reflected from the top polymerized surface, within the focal volume. It is to be emphasized that the structures inside polymer could be written flawlessly even at extremely high speed of 100mm/s or more, which, is the highest ever reported writing speed using this technique [32]. Complex 2D/3D micro/nano structures containing round or geometric pattern have been written using this method. Figure 3(c) shows sample SEM image of highly complex 2D structure in form of official logo of our institute "Indian Institute of Technology, Bombay". The dimension of this logo is 280μm x 280μm and is written in pure resin at a scanning speed of 1500μm/s using laser power at 0.18W at 800nm wavelength. Figure 3(d) is the zoomed image of 3(c) showing connected structures of submicron and nano feature sizes. At constant laser power, the feature size was observed to vary with scanning speed (data not shown here).

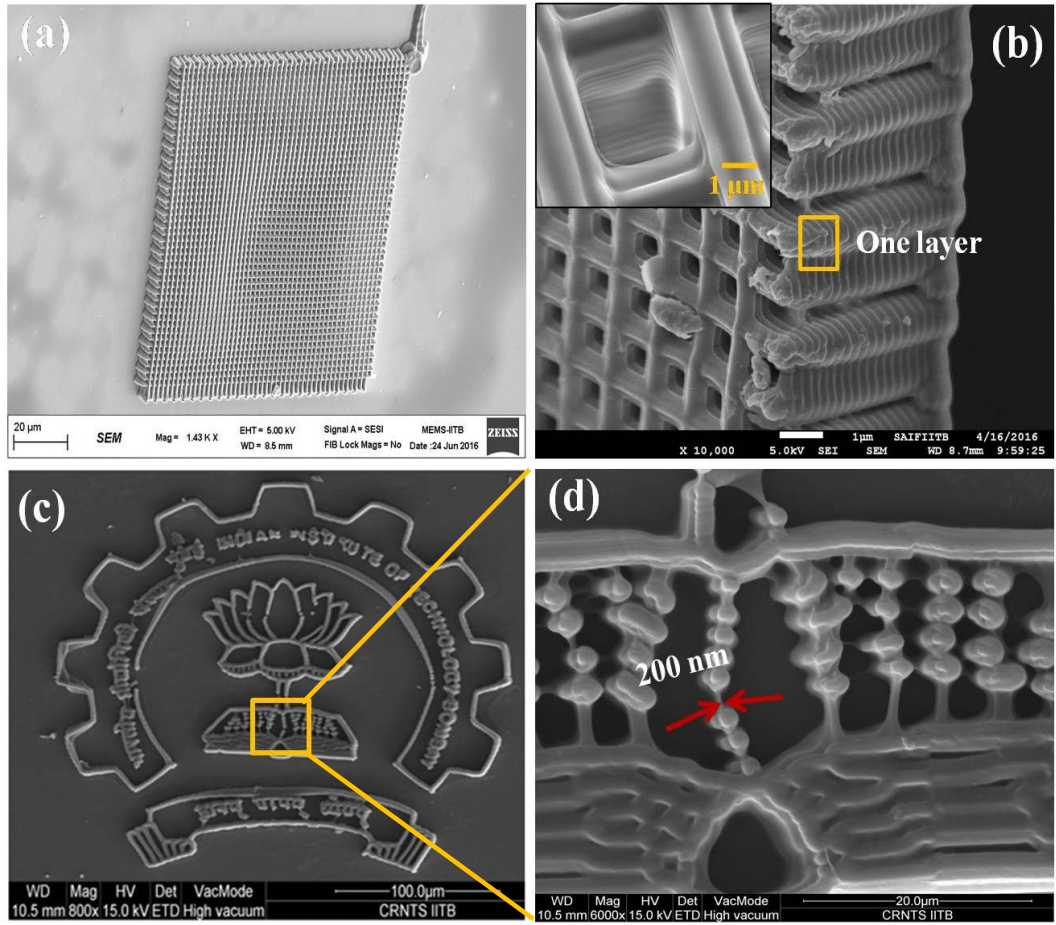

*Fig. 3. (a) 3D mesh of 6 layers written at 0.25W laser power and 5000μm/s scanning speed. (b) Shows the cross section view of the mesh structure where single layer has been marked in yellow box. Due to standing wave formation within the focal volume single line shows 3 edges. Inset in 3(b) is the zoomed in image of single mesh showing multiple layers. Figure 3(c) is SEM image of our institute "Indian Institute of Technology Bombay" logo, written at 1500μm/s scanning speed and 0.18W laser power. Figure 3(d) shows the fine polymer lines ~ 200 nm connecting broadly polymerised spots.*

The exact mechanism of sub-wavelength writing with high amount of free radical photoinitiator is under investigation. Photopolymerization using two photon dyes involving linear or non-linear process has been well explored in the past. Nonlinear interaction is achieved by using two-photon absorbing dyes such as alexa fluor, malachite green etc, followed by transferring of energy to the curing agent or photo initiator present in photo-polymer matrix. Then standard photo-polymerization starts with the absorption of single photon of specific wavelengths (usually *uv* photon) in a small focal volume of resin. The photoinitiator transforms into a reactive specie upon exposure, which in turn reacts with the monomer to produce monomer radicals. This monomer radical combines with other monomers and starts a chain reaction of polymerization[7] and can be expressed as;

Two photon initiation of Lucirin TPOL giving rise to two free radicals upon light absorption($R^*$),

$$I \xrightarrow{2h\vartheta} 2R^*$$

Free radical polymerization initiation of first monomer;

$$R^* + M \rightarrow R-M^*$$

Chain propagation with second monomer;

$$R-M^*_n + M \rightarrow R-M^*_{n+1}$$

Termination by free radical combination;

$$2R^* + M_n \rightarrow R-M_{2n}-R$$

Two-photon absorption processes utilize simultaneous absorption of two-photon to reach the same energy level as of single high energy photon and start the chain reaction. Since this is a non-linear process, reaction takes place in a very small volume where intensity of the beam is very high. Above the threshold value, femtosecond laser pulses with extremely high peak intensities leads to a spatially confined absorption volume in the laser focus, hence triggering photopolymerization of a volume pixel ("voxel") of typically ellipsoidal shape. This voxel is scalable in size depending on the irradiated laser intensity and focusing optics, allowing for features sizes down to 140 nm, i.e., much smaller than the irradiated wavelength of light. Typically two-photon absorption cross-section of photoinitiators is low, thus high laser threshold is required for starting photopolymerization reaction. Standard concentration of photointiator in monomer/oligomer matrix is fixed to few percent. In this work, we have used 20 weight% photoinitiator to achieve sub-wavelength line widths. By increasing the photoinitiator concentration, polymerization reactions starts simultaneously in entire exposed area and terminates by reacting with the radical end in the vicinity, giving rise to high resolution features.

## 4. Conclusion:

In a nutshell, we present here an economical high throughput fabrication method using two-photon lithography technique. This method has been used to fabricate complex 2D/3D micro/nano structures with sub-wavelength resolution inside polymer matrix without using any expensive two-photon dye. Feature size with line widths as small as 140nm has been demonstrated for sartomer photopolymer with 20 weight% loading of photoinitiator. The impact of critical process parameters such as laser power and scan speed, on feature size of the structure has been studied. This approach is expected to pave way for direct, inexpensive fabrication of photonic materials which holds potential application towards 3D metamaterials, plasmonics and other optical technologies.

**Funding & Acknowledgements:**

This work was supported by the Department of Science and Technology, Solar Energy Research Initiative (SERI), Government of India grant via sanction order no. DST/TM/SERI/2k10/12/(G) and the Industrial Research and Consultancy Services, Indian Institute of Technology Bombay, grant no. 11IRCCSG025.